\documentstyle[12pt]{article}
\newcommand\beq{\begin{equation}}
\newcommand\eeq{\end{equation}}
\begin{document}

\begin{center}
QCD PHASE TRANSITION IN HOT HADRONIC MATTER
\renewcommand{\thefootnote}{\fnsymbol{footnote}}
\footnote{Invited talk at ISMEP 94 (International Symposium on
Medium Energy Physics), Beijing 1994}
\end{center}

\vspace{1.5cm}

\begin{center}
H.J. PIRNER and B.-J. SCHAEFER\\
{\it Inst. f. Theoret. Physik, Univ. Heidelberg,\\
Philosophenweg 19,
D-69120 Heidelberg, Germany}
\end{center}

\vspace{1.0cm}

{\small
\begin{center}
ABSTRACT\end{center}\begin{quotation}
We analyse the QCD chiral phase transition in the nonlinear and linear
$\sigma$-model.
The strategy is the same in both cases. We fix the parameters of the effective
meson
theory at temperature $T=0$  and  extrapolate the models to   temperatures in
the
vicinity of the phase transition. The linear $\sigma$-model in $SU(3)\times
SU(3)$
gives a crossover around $T_c\approx190$ MeV. Around this temperature chiral
$SU(2)\times SU(2)$ is almost restored. We also calculate meson masses as a
function of temperature.\end{quotation}}

\vspace{0.5cm}

\setlength{\baselineskip}{14pt}
\noindent {\bf1. Introduction}

\vspace{0.5cm}

Relativistic nucleus-nucleus collisions with cm--energies $E_{cm}\geq20$
GeV/nucleon
create
large systems of sizes $R>20$ fm at freeze-out with
$>10^4$ pions \cite{schu}.It is natural to try statistical methods to describe
such hadronic fireballs. A good starting point may be to use equilibrium
thermodynamics with pions when one is interested in the later stages of
these collisions, where the temperature is below possible phase transitions.
The low energy interaction of pions is fully determined by chiral
symmetry \cite{gerb},\cite{lee}.
Below $T \approx 120$ MeV this interaction can be parametrized by the nonlinear
sigma
 model $O(4)$  with
one scalar ($\sigma$) and three pseudoscalar $\vec{\pi}$-fields, which
are constrained by the condition $\sigma^2+\vec{\pi}^2=f_\pi^2$, where
f$_\pi$ is the pion decay constant.
%%%%%%%%%%%%%%%%%%%%%%%%%%%%%%%%%%%%%%%%%%%%
% M
%%%%%%%%%%%%%%%%%%%%%%%%%%%%%%%%%%%%
Above this temperature region also heavier hadrons give
a
non-negligible contribution to the condensates and thermodynamic
quantities.  One way of including part of the heavier mesons is
provided by the choice
of $SU(3)\times SU(3)$ as chiral symmetry group rather than
$SU(2)\times SU(2)$. The linear
$SU(3)\times SU(3)$
sigma model includes a nonet of pseudoscalar  $(O^-)$-fields and a
nonet of
scalar $(O^+)$-fields. The spontaneous breaking of the
$SU(3)\times SU(3)$ symmetry leads to
massless $(O^-)$ Goldstone modes. Obviously a massless
pseudoscalar octet does
not provide an adequate approximation to the experimentally
observed meson spectrum.
Therefore we include explicit symmetry breaking terms to account
for the physical
mass values of the octet-fields. A determinant term in the meson fields
guarantees
the correct mass splitting of the $\eta-\eta'$ masses which is due to
the
$U(1)$-anomaly. It reflects  the 't Hooft-determinant on the quark
level.

%%%%%%%%%%%%%%%%%%%%%%%%%%%%%%%%%%%%%%%%%%%%

The experimental challenge is to measure the equation of state of pions from
the inclusive pion spectra. The theoretical task is to calculate this
equation of state. For this purpose we need reliable techniques to treat
field theories at finite temperature. One can then extrapolate from the
measured physics at $T=0$ to the yet unknown physics at high temperatures.
A very accurate treatment of the soft modes with lowest mass is essential
at low temperatures. We calculate the partition function $\cal Z$ in terms of
a selfconsistent field which is chosen to extremize $\ln {\cal Z}$. It gives
effective
masses to the meson fields. This saddle point approximation to the partition
function
corresponds to the leading order of a $1/N$ expansion.
%%%%%%%%%%%%%%%%%%%%%%%%%%%%%%%%%%%%%%%%%%%%%%%%%% % Neuer Teil
%%%%%%%%%%%%%%%%%%%%%%%%

The paper is organized as follows. In section 2 we discuss the nonlinear
$\sigma$-model for $SU(2) \times SU(2)$. In section 3 we calculate the
partition
function in the linear $\sigma$-model for $SU(3)\times SU(3)$. Section 4 is
devoted to
a short discussion.

\vspace{0.5cm}

\noindent{\bf2. The Nonlinear $\sigma$-Model: $SU(2)\times SU(2)$}

\vspace{0.5cm}

The partition function $\cal Z$ for the $SU(2)\times SU(2)$ nonlinear
$\sigma$-model is
given in terms of the $O(4)$-multiplet $(n_0,\vec{n}) = (\sigma,\vec{\pi})$
with
$n^2=\sigma^2+\vec{\pi}^2$ as:
\beq
 {\cal Z} = \int{{\cal D}n(x)} \prod_{x} \delta(n^2(x)-f_{\pi}^2)
\exp \{-\int_0^{\beta}d\tau \int_V d^3x
[\frac{1}{2}(\partial_{\mu}n)^2-cn_0] \}\;.
\eeq
At zero temperature $T:=1/\beta=0$ the parameters of the model are well
known.The
pion decay constant $f_{\pi}$ equals to 93 MeV. The classical vacuum
expectation value
of $n_0$ is determined as $<n_0>=f_{\pi}$ by minimizing the vacuum energy.
Expanding the dependent field $n_0=\sqrt{f_{\pi}^2-\vec{n}^2}$ to leading order
in $\vec{n}^2/f^2_\pi$, one obtains the mass of the pion as
$m_{\pi}^2=c/f_{\pi}$.

The basic idea of our method is to eliminate the nonlinear
constraint $n^2=\sigma^2+\vec{\pi}^2=f_{\pi}^2$ by the introduction of an
auxiliary field $\lambda(x)$ via an integral along the imaginary axis
\beq
{\cal Z} = \int\limits_{a-i\infty}^{a+i\infty}{{\cal D}\lambda(x)\int {\cal
D}n(x)}
\exp\{-\int_0^{\beta}d\tau \int_V d^3x
[\frac{1}{2}(\partial_{\mu}n)^2+\lambda(n^2-f_{\pi}^2)-cn_0] \}\;.
\eeq
After shifting the zeroth component $n_0$ to $\tilde{n}_0$
\begin{eqnarray*}
\tilde{n}=(\tilde{n}_0,\tilde{\vec{n}})=(n_0-\frac{c}{2\lambda},\vec{n})\;,
\end{eqnarray*}
we obtain a
Gaussian action for the $O(N)$ multiplet field $\tilde{n}$, when we evaluate
Eq.(2) in a saddle point approximation.
The resulting partition function is given as
\beq
{\cal Z} = \int{\cal D}\tilde{n}(x)
\exp\{-\int_0^{\beta}d\tau \int_V d^3x
[\frac{1}{2}(\partial_{\mu}\tilde{n})^2+\lambda\tilde{n}^2-\lambda f_{\pi}^2
-\frac{c^2}{4\lambda}]\}\;.
\eeq
Here we have dropped the  $\lambda$-integration and chosen
$\lambda(x)=\lambda=const$.
The optimal choice for $\lambda$ will be determined later, cf.~eq.~(11).

Upon Gaussian integration over the four ($N=4$) $\tilde{n}$-fields we end up
with a
partition function of a free relativistic Bose gas with $N=4$ components
and effective masses
\beq
m_{eff}^2= 2\lambda.
\eeq
\beq
{\cal Z} = \exp \{ -\beta V[U_{T}(m_{eff}^2,\Lambda)+U_0(m_{eff}^2,\Lambda)
-\lambda f_{\pi}^2-\frac{c^2}{4\lambda} ] \}\;.
\eeq
Here $U_T$ denotes the contribution from thermal fluctuations
\beq
U_{T}(m_{eff}^2,\Lambda) = 4 \frac{1}{\beta} \int_0^{\Lambda}
\frac{d^3 k}{(2\pi)^3} \ln ( 1 - e^{-\beta \sqrt{\vec{k}^2+m_{eff}^2}} )
\eeq
and $U_0$ the zero point energy
\begin{eqnarray}
U_0(m_{eff}^2,\Lambda) & = & 4 \int_0^\Lambda \frac{d^3 k}{(2\pi)^3}
\frac{1}{2}\sqrt{\vec{k}^2+m_{eff}^2}\;.
\end{eqnarray}
We regularize the $k$--integrations in Eqs.~$(7)$ and $(8)$ with a cut-off
$\Lambda$,
since we do not believe our $\vec{\pi}$--effective theory to be correct for
momenta beyond $\Lambda$. At momenta $k>\Lambda$ the compositness of the pions
manifests itself in resonance excitations and/or higher derivative
couplings of the pion states, which are neglected. For the numerical
calculations we take cut--off values $\Lambda = 700$ MeV, $800$ MeV, $1000$
MeV.

After regularization we adopt the following renormalization procedure.
We define a renormalized potential at arbitrary $T$ according to
\begin{eqnarray}
U^{ren}(m_{eff}^2,\Lambda) & = & U_T(m_{eff}^2,\Lambda)
+ U_0(m_{eff}^2,\Lambda)\nonumber\\
                   & & - [ U_0(m_{\pi}^2,\Lambda)+(m_{eff}^2-m_{\pi}^2)
\left. \frac{\partial U_0}{\partial m_{eff}^2} \right|_{m_{\pi}^2} ].
\end{eqnarray}
The two subtractions guarantee the two renormalization
conditions at $T=0$
\begin{eqnarray}
m_{eff}^2(\lambda_0) & = & m_{\pi}^2\\
<n_0> & = & f_{\pi}\;.
\end{eqnarray}

It is well known that the nonlinear sigma model is not renormalizable in four
dimensions.  Therefore higher order divergences can only be compensated by
higher order
derivative terms in the original action.
The coefficients of these higher order terms have to be determined by
experiment. We do not include such terms in contrast to ref.~[2]. In section 3
we will extend the calculation to the linear $\sigma$--model $(SU(3)
\times SU(3))$ which contains higher masses and strange mesons.

The thermodynamic observables at finite $T$ are obtained from the partition
function $\cal Z$ approximated as
\beq
{\cal {Z}}(\Lambda,T) = \exp \left\{ -\beta V \left[ U^{ren}(m_{eff}^2
(\lambda^{ \ast}),\Lambda)-\lambda^{ \ast} f_{\pi}^2 -
\frac{c^2}{4\lambda^\ast}
\right] \right\}\;,
\eeq
where $\lambda^\ast(T)$ extremizes $\ln{ \cal Z}$ at a given temperature
$T\neq0$.
The saddle point equation
for $\lambda^\ast$ is solved numerically, since in the interesting temperature
range the relevant parameter $m_{eff}/T = \sqrt{2\lambda^\ast}/T$ can
have   values in the range $0\leq m_{eff}/T < \infty$.

Let us first study, how the order parameter of chiral symmetry
breaking $<n_0>$ behaves as a function of temperature. In Fig.~1 we present
the result for $\frac{<n_0(T)>}{<n_0(T=0)>} =
\left. \frac{\partial \ln {\cal Z}(T)}{\partial c} \right/
\frac{\partial \ln {\cal Z}(T=0)}{\partial c}$.
%\begin{figure}[tbp]
%\vspace{10.0cm}
%\caption{
%The order parameter for chiral symmetry breaking
%$<{\bar q}q(T)>/<{\bar q}q(T=0)>$ for two different cut-offs $\Lambda=700$
%MeV (solid line) and $\Lambda=1000$ MeV (dashed line).
%The diamonds represent the result of the linear $\sigma$-model in section 3.
%The crosses are the result of ref.~2.}
%\end{figure}
In quark language this ratio corresponds to the ratio of the quark
 condensate $<{\bar q}q(T)>$ at finite temperature over the quark condensate
at $T=0$, since the symmetry breaking term of the $O(4)$ Lagrangian
${\cal L}_{SB}= cn_0$ is identified with  the symmetry breaking
 term ${\cal L}_{SB} = -2m\bar{q}q$
in the QCD--Lagrangian.
We also show the result of the linear $\sigma$-model $SU(3)\times SU(3)$,
which are presented in the next section, and the results of chiral perturbation
theory
\cite{gerb}. The result for$<\bar{q}q(T)>/<\bar{q}q(T=0)>$ is rather
insensitive to the
cut-off. It agrees well with the three loop calculation of ref.~[2].
Chiral symmetry is only very gradually restored. At low temperature the
$\pi\pi$--interaction is weak and $<\bar{q}q>$ does not change very much.

\vspace{0.5cm}

\noindent{\bf3. The Linear $\sigma$-Model: $SU(3)\times SU(3)$}

\vspace{0.5cm}

For a Euclidean metric the Lagrangian of the linear sigma--model is
given
as
\begin{eqnarray}
\label{eq:1}
{\cal L} & = & \frac{1}{2}{\rm tr}\partial_{\mu}\Phi\partial_{\mu}\Phi^+
-\frac{1}{2}\mu_{0}^{2} {\rm tr}\Phi\Phi^++f_{1}\left(
{\rm tr}\Phi\Phi^+
\right)^{2}+f_{2}{\rm tr}\left(\Phi\Phi
^+\right)^2\nonumber\\
 & & +g\left(\det\Phi+\det\Phi^+\right)-\varepsilon_{0}\sigma_{0}-
\varepsilon_8\sigma_{8},
\end{eqnarray}
where the ($3\times 3$) matrix field $\Phi(x)$ is given in terms of
Gell--Mann
matrices $\lambda_\ell\  (\ell=0,\ldots ,8$) as

\begin{equation}
\Phi = \frac{1}{\sqrt{2}} \sum_{\ell = 0}^{8} \left( \sigma_{\ell}+i
\pi_{\ell}
\right)\lambda_{\ell}.
\end{equation}
Here $\sigma_{\ell}$ and $\pi_{\ell}$ denote the nonets of scalar
and pseudoscalar
mesons, respectively. As order parameters for the chiral transition
we choose the meson condensates $\protect{<\sigma_0>}$ and $<\sigma_8>$.
The chiral symmetry of
${\cal L}$  is explicitly broken by the term
$(-\varepsilon_0\sigma_0-\varepsilon_8\sigma_8)$, corresponding
to the finite quark
mass term $2m_q\bar q q+m_s \bar s s$ on the quark level. The
chiral limit is realized
for vanishing external fields $\varepsilon_0$ and $\varepsilon_8$.
Note that the
action $S=\int d^3x d\tau{\cal L}$ with ${\cal L}$ of Eq.~(\ref{eq:1})
may be
regarded as an effective action for QCD, constructed in terms of an
order parameter
field $\Phi$ for the chiral transition. It plays a similar role to
Landau's free
energy functional for a scalar  order parameter field for
investigating the phase
structure of a $\Phi^4$-theory\cite{pis}.

The six unknown couplings of the sigma-model (Eq.~(\ref{eq:1}))\
$(\mu^2_0, f_1, f_2,
g,\varepsilon_0,\varepsilon_8)$ are assumed to be temperature
independent and adjusted
to the pseudoscalar masses  at zero temperature. Further
experimental input parameters
are the pion decay constant  $f_\pi=94$ MeV and a high lying
$(O^+)$  scalar mass
$m_{\sigma_\eta}=1.20$ GeV (cf. Table 1). For the remaining scalar masses
and the coupling
constants we obtain the values of Table 1.

\vspace{0.5cm}

\begin{table}[htb]
\begin{center}
\caption{\label{values}Tree level parametrization of the
$SU(3)\times SU(3)$ linear
sigma model.}
\begin{tabular}{|l|l|l|l|l|l|}
\multicolumn{6}{c}{Input} \\
\hline
$m_{\pi}$ [MeV]& $m_{K} [MeV] $ & $m_{\eta} [MeV]$ &
$m_{\eta'}$ [MeV]&
$f_{\pi}$ [MeV] & $m_{\sigma_\eta}=m_{f_{0}}$ [MeV]\\
\hline
130. & 491. & 544. & 1046. & 94. & 1200.\\
\hline
\multicolumn{6}{c}{} \\
\multicolumn{6}{c}{Output} \\
\hline
$\mu_{0}^{2}$ [GeV$^{2}$] & $f_{1}$ & $f_{2}$ & $g$ [GeV] &
$\varepsilon_{0}$
[GeV$^{3}$] & $\varepsilon_{8}$ [GeV$^{3}$]\\
\hline
0.059 & 4.17 & 4.48 & -1.8 & 0.026 & -0.035\\
\hline
\multicolumn{6}{c}{}\\
\hline
$m_{a_{0}}$ [GeV] & $m_{K_{0}^{\ast}}$ [GeV] & $m_{f_{0}'}$
[GeV] & & & \\
\hline
1.0116 & 1.0312 & 0.7495 & & & \\
\hline
\end{tabular}
\end{center}
\end{table}
The interpretation of the observed scalar mesons is controversial.
There are
good reasons to interpret the ($0^+$) mesons at 980 MeV
as meson bound states. The model underestimates the
strange quark mass
splitting in the scalar meson sector, the value for $m_{K^*_0}$
comes out too small.

The effective theory can be related to the underlying QCD
Lagrangian by comparing
the symmetry breaking terms in both Lagrangians and identifying
terms with the same
transformation behaviour under
$SU(3)\times SU(3)$. Taking expectation values in these equations
we obtain the following relations between the light quark
condensates, strange quark
condensates and meson condensates

\begin{eqnarray}
\label{eq:3}
\varepsilon_{0}\left<\sigma_{0}\right> & = &
-\frac{1}{3}\left(2\hat{m}+m_{s}\right)\left(2\left<
\bar{q}q\right>+\left<\bar{s}s\right>\right)\nonumber\\
\varepsilon_{8}\left<\sigma_{8}\right> & = &
-\frac{2}{3}\left(\hat{m}-m_{s}\right)\left(2\left<
\bar{q}q\right>-\left<\bar{s}s\right>\right).
\end{eqnarray}
We use $\hat m\equiv (m_u +m_d)/2
=(11.25\pm 1.45)$ MeV and $m_s=(205\pm 50)$
MeV for the light and strange quark masses at a scale
$\Lambda=1$
GeV~\cite{nar}.  From
the scalar meson condensates at $T=0$,
$\left<\sigma_{0}\right>=0.14$
GeV and $\left<\sigma_{8}\right>=-0.03$ GeV we get

\begin{eqnarray}
\left<\bar{q}q\right> & = & -\left(243.8\pm 60 {\rm
MeV}\right)^{3}\nonumber\\
\left<\bar{s}s\right> & = & -\left(285.0\pm 30 {\rm MeV}\right)^{3}
\end{eqnarray}
in accordance with values  from PCAC relations~\cite{nar} within
the
error bars. Since we treat the coefficients
$\varepsilon_0,\varepsilon_8$ of
$<\sigma_0>$ and $<\sigma_8>$, and $\hat m, m_s$ of $<\bar q q>$
and $<\bar s s>$
as temperature independent, we will use Eqs.~(\ref{eq:3}) for all
temperatures
to translate our results for meson condensates into quark
condensates.

We also check that the pseudoscalar meson mass squares, in
particular $m^2_\pi$ and
$m^2_K$ are linear functions of the symmetry breaking parameters
$\varepsilon_0,\varepsilon_8$. Varying
$\varepsilon_0,\varepsilon_8$ while keeping
the other couplings fixed we can simulate the sigma model at
unphysical meson
masses. Since the current quark masses are assumed to depend
linearly on
$\varepsilon_0$ and $\varepsilon_8$, an arbitrary meson  mass set
can be related to
a mass point in the $(m_{u,d},m_s)$-plane by specifying the choice
of
$(\varepsilon_0,\varepsilon_8)$. This may be useful in order to
compare our results
for meson (and quark) condensates with lattice simulations of the
chiral transition.

The thermodynamics of the linear sigma model is determined by
the partition
function with the Lagrangian of Eq.~(\ref{eq:1})

\begin{equation}     \label{eq:5}
Z=\int{\cal D}\Phi \exp\left\lbrace-\int^\beta_0 d\tau\int d^3
x{\cal L}
(\Phi(\vec x, \tau))\right\rbrace.
\end{equation}
We will treat $Z$ again in a saddle point approximation. As mentioned above,
the saddle point
approximation
amounts to the leading order of a $1/N$-expansion. In this model
$N=2 N^2_f=18$. Note that $\cal L$ of Eq.~(\ref{eq:1}) would be
$O(N)$-invariant, if
$f_2=0$ and $g=0$. Our input parameters lead to non-vanishing
values
of $f_2$ and $g$, therefore the $O(N)$-symmetry is only
approximately realized.

We calculate the effective potential as a constrained free energy
density $U_{eff}
(\xi_0,\xi_8)$, that is the free energy density of the system under
the constraint
that the average values of $\sigma_0$ and $\sigma_8$ take some
prescribed values
$\xi_0$ and $\xi_8$.  The values $\xi_{0_{min}}$ and
$\xi_{8_{min}}$ that
minimize $U_{eff}$, give the physically relevant, temperature
dependent  vacuum
expectation values, i.e. $<\sigma_0>=\xi_{0_{min}},\
<\sigma_8>=\xi_{8_{min}}$.
Hence we start with the background field ansatz
\begin{eqnarray}\label{eq:6}
\sigma_0&=& \xi_0+\sigma'_0\nonumber\\
\sigma_8&=& \xi_8+\sigma'_8,
\end{eqnarray}
where $\sigma'_0$ and $\sigma'_8$ denote the fluctuations around
the classical
background fields $\xi_0$ and $\xi_8$. All other field components
are assumed to have
zero vacuum expectation value, i.e. $\sigma_\ell=\sigma'_\ell$ for
$\ell=1,\ldots,7$
and $\pi_\ell=\pi'_\ell$ for $\ell=0,\ldots,8$.
The relation between the effective potential $U_{eff}$ and $Z$ is
given by
\begin{eqnarray}
\label{eq:7}
Z&=&\int d\xi_0\int d\xi_8  e^{-\beta V
U_{eff}(\xi_0,\xi_8)}\;.
\end{eqnarray}

Next we insert the background field ansatz (\ref{eq:6}) in $\cal L$
and expand the
Lagrangian in powers of $\Phi'=\frac{1}{\sqrt
2}\sum^8_{\ell=0}(\sigma'_\ell
+i\pi'_\ell)\lambda_\ell$. The constant terms in $\Phi'$ lead to the
classical part
of the effective potential $U_{class}$.
Linear terms in $\Phi'_\ell$ vanish for all $\ell=0,\ldots,8$ due to
the
$\delta$-constraints in Eq.~(\ref{eq:7}).
Quadratic terms in $\Phi'$ define the isospin multiplet masses
$m^2_Q$, where
$Q=1,\ldots,8$ labels the multiplets.

The cubic part in $\Phi'$ will be neglected, while the quartic term
${\cal L}^{(4)} (\Phi')$
is quadratized by introducing an auxiliary matrix field $\sum(\vec
x,\tau)$.
This is a matrix version of a Hubbard-Stratonovich transformation
\cite{hub}.

In the saddle point approximation we drop $\int{\cal D}\Sigma$ and use a
$SU(3)$-symmetric
ansatz with a diagonal matrix $\sum = \mbox{diag}(s,s,s)$.
Hence the effect of the quadratization procedure is to induce an
extra mass term
$(s+\mu^2_0)$ and a contribution $U_{\rm saddle}$ to $U_{eff}$,
which is independent
of $\xi_0$ and $\xi_8$.
This way we finally end up with the following expression for $\hat
Z$

\begin{eqnarray}
\label{eq:17}
\hat Z(\xi_0,\xi_8,s)&=&e^{-\beta V(U_{\rm class}+U_{\rm
saddle})}\cdot\nonumber\\
& &\cdot\int\prod^8_{Q=1}{\cal D}\varphi'_Q e^{-\int^\beta_0
d\tau\int d^3 x\frac{1}{2}\sum_Q g
(Q)(\partial_\mu\varphi'_Q\partial_\mu
\varphi^{\prime\dagger}_Q + X^2_Q \varphi'_Q\varphi^{\prime\dagger}_Q)}
\end{eqnarray}
where
\begin{equation}\label{eq:18}
X^2_Q\equiv m^2_Q+\mu^2_0+s\;,
\end{equation}
$\varphi'_Q$ denotes $\sigma'_Q$ for $Q=1,\ldots,4$ and
$\pi'_Q$
for $Q=5,\ldots,8,\ g(Q)$ is the multiplicity of the isospin multiplet.
We have
$g(1)=3$ for the pions, $g(2)=4$ for the kaons, $g(3)=1=g(4)$ for
$\eta,\eta'$,
respectively. Correspondingly, the multiplicities for the scalar
nonets are $g(5)=3,
\ g(6)=4,\ g(7)=1,\ g(8)=1$ for the $a_0,K^*_0, f_0,f'_0$-mesons.

Thus we are left with an effectively free field theory. The only
remnant of the
interaction appears in the effective mass squared $X^2_Q$ via the
auxiliary field
$s$.

The choice of a self-consistent effective meson  mass squared has
been pursued
already in Refs.~\cite{frei,mey}.  This is an essentially new
ingredient compared to
earlier calculations of the chiral transition in the linear sigma model
\cite{gol}.
The positive contribution of $s$ to the effective mass extends the
temperature region,
where imaginary parts in the effective potential can be avoided. In
general,
imaginary parts are encountered, when the effective mass
arguments of logarithmic
terms become negative. They are an artifact of the perturbative
evaluation of the
effective potential and of no physical significance, as long as the
volume is infinite.
In our application the optimized choice for $s$ will increase as
function of
temperature and lead to positive $X^2_Q$ over a wide range of
parameters.

Gaussian integration over the fluctuating fields $\Phi'$ in
Eq.~(\ref{eq:17}) gives
\begin{eqnarray}\label{eq:19}
\hat Z(\xi_0,\xi_8,s)&=&\exp\left\lbrace-\beta V\left[ U_{\rm
class}+U_{\rm saddle}+
\right.\right.\nonumber\\
& & +\frac{1}{2\beta}\sum^8_{Q=1}g(Q)\sum_{n\in
Z}\int\frac{d^2k}{(2\pi)^3}
\left.\left.\ln\left(\beta^2 (\omega^2_n + \omega^2_Q ) \right)\right]
\right\rbrace
\end{eqnarray}
where
\begin{equation}\label{eq:20}
\omega^2_Q\equiv k^2+X^2_Q,
\end{equation}
and
\begin{equation}
\label{eq:21}
\omega^2_n\equiv(2\pi n/T)^2
\end{equation}
denote the Matsubara frequencies. In contrast to our former
approach \cite{mey}
we keep all Matsubara frequencies and evaluate $\sum_{n\in
Z}$ in the
standard way, see e.g. \cite{kap}. The result is
\begin{eqnarray}
\hat Z(\xi_0,\xi_8;s)&=&e^{-\beta V U_{\rm
eff}(\xi_0,\xi_8;s)}\label{eq:22}\\
U_{\rm eff}(\xi_0,\xi_8;s)&=& U_{\rm class}+U_{\rm saddle}
+U_{\rm th}
\label{eq:23}\\
U_{\rm th} & \equiv & \frac{1}{\beta}\sum_{Q=1}^{8}g(Q)\int
\frac{d^{3}k}
{(2\pi)^{3}}\ln\left(1-e^{ -\beta\omega_Q}\right)\label{eq:24}
\end{eqnarray}

Here we have indicated that $\hat Z$ and $U_{\rm eff}$
still depend explicitly on the auxiliary field $s$.

The linear sigma model is a renormalizable theory, and in principle the zero
point
energy can be calculated and renormalized.  We do, however,
believe that this model is an effective description for QCD already at tree
level.

Now we are prepared to determine the temperature dependence of
the  order parameters
$<\xi_0>(T),\ <\xi_8>(T)$  from the minima of $U_{\rm
eff} (\xi_0,\xi_8;
s^*)$. Thermodynamic quantities like energy densities, entropy densities
and pressure
can be derived from $Z$ in the standard way, if $Z$ is approximated
as
\begin{equation}\label{eq:30}
\hat Z \equiv e^{-\beta V U_{\rm
eff}(\xi_0,\xi_8;s^*)}.
\end{equation}

For the parameters of Table~\ref{values} we vary the temperature
and determine
for each $T$ the extremum of $U_{\rm eff}$ as a function of
$\xi_{0}$,
$\xi_{8}$ and $s$. The extremum is a minimum with respect to
$\xi_{0}$ and $\xi_{8}$ and a maximum with
respect to  $s$. For the search of the saddle point it is necessary to continue
the
effective potential into the region of complex effective masses $X^2_Q$ (cf.
Ref.~\cite{meyo}).

%\begin{figure}[tbp]
%\vspace{10.5cm}
%\caption{ Normalized light quark condensate
%$\frac{\left<\bar{q}q\right>}{\left<\bar{q}q
%\right>_{T=0}}$ vs temperature (full curve) and
%normalized strange quark condensate
%$\frac{\left<\bar{s}s\right>}{\left<\bar{s}s
%\right>_{T=0}}$ vs temperature (dashed curve).}
%\end{figure}
In Fig.~2  we show the variations of $\frac{<\bar q
q>(T)}{<\bar q q>T=0}$
and $\frac{<\bar s s>(T)}{<\bar s s>T=0}$ as a function of
temperature obtained from
$<\xi_0>(T)$ and $<\xi_8>(T)$ with the help of Eq.~(\ref{eq:3}). We
observe a
gradual decrease of the light quark condensate, whereas the
strange quark condensate
stays almost constant.

In our lowest order calculation  the temperature
dependence of these masses is determined by the temperature
dependence of the
condensates~[cf. Fig.~3].
%\begin{figure}[tbp]
%\vspace{10.5cm}
%\caption{ Meson masses in GeV as a function of temperature
%$T$.} \end{figure}
The masses $m^2_\pi$ and $m^2_{\sigma_{\eta '}}$ become
degenerate after the crossover. The $\pi-K$-
splitting is increased
rather than reduced. Accordingly the strange meson contribution to
the energy density
in this temperature region is reduced compared to the low-
temperature hadron gas.

In Fig.~4 we give the energy density $u/T^4$, entropy density $s/T^3$ and
pressure $p/T^4$.
%\begin{figure}[tbp]
%\vspace{10.5cm}
%\caption{ Normalized energy density
%$\frac{u}{T^{4}}$ , entropy density $\frac{s}{T^3}$ and pressure
%$\frac{p}{T^{4}}$
%vs temperature.}
%\end{figure}
Sizeable  contributions to
$u$ come mainly
from 8 degrees of freedom, the pions, the kaons
 and the $f'_0$ meson.

\vspace{0.5cm}
\newpage
\noindent{\bf 4. Discussion of the Results}

\vspace{0.5cm}

For low temperatures the physics of the nonlinear $SU(2)\times SU(2)$ and
linear $SU(3)\times SU(3)$ $\sigma$-model are identical. In Fig.~1 we show the
light
quark condensates calculated in both models. Above $T \approx 120$ MeV the
extra
degrees of freedom in the $SU(3)\times SU(3)$ calculation become important. At
higher temperatures $T \gg T_0\approx190$~MeV also the linear sigma model will
certainly fail as an effective model for QCD due to the lack of quark-gluon
degrees of
freedom. Nevertheless it would be interesting to study, at what temperature
 the full $SU(3)\times SU(3)$ symmetry is restored. At very high
temperatures the
effective potential becomes proportional to $\sum_Q X^2 (Q) T^2$,
the linear terms proportional to $\sigma_0$ and $\sigma_8$
in the masses of $O^+$ and $O^-$ mesons cancel and
temperature tries to
fully restore the broken symmetry.

Finally we remark that
the crossover occurs around $T_0\approx190$~MeV, which
is rather close to the Hagedorn temperature
$T_H
(T_H\sim160$~MeV). This may not be entirely accidental.
In our model the
$1/N$-expansion means a large   number of  flavours, since
$N=2\cdot N^2_f$.
In order to keep QCD an asymptotically free theory also  the
number of colours $N_c$
has to increase. Correspondingly our approximation is similar to
Hagedorn's
description of the hadron gas as a resonance gas. We expect that corrections
from
subleading
terms in our $1/N_f$-expansion will implicitly amount to
corrections also to the
large $N_c$-limit.
The chiral transition for unphysical values of strange quark and
light quark masses
will be investigated in a future publication \cite{meyo}.

\section*{\bf {\Large {\bf Figure Captions}}}

\newcounter{fig}
\begin{list}{\bf Fig. \arabic{fig}:}{\usecounter{fig}
\leftmargin1.6cm
\labelwidth1.6cm
\labelsep0.2cm }

\item The order parameter for chiral symmetry breaking
$\left<{\bar q}q(T)\right>/\left<{\bar q}q(T=0)\right>$
for two different cut-offs $\Lambda=700$
MeV (solid line) and $\Lambda=1000$ MeV (dashed line).
The diamonds represent the result of the linear $\sigma$-model in section 3.
The crosses are the result of ref.~2.
\item
Normalized light quark condensate
$\left<\bar{q}q\right>/\left<\bar{q}q
\right>_{T=0}$ vs temperature (full curve) and
normalized strange quark condensate
$\left<\bar{s}s\right>/\left<\bar{s}s
\right>_{T=0}$ vs temperature (dash\-ed curve).
\item Meson masses $X_Q$, $Q=1,\ldots,8$ as a function of temperature $T$.
 ($X_{\sigma_\pi} = m_{a_0}$~, $X_{\sigma_K} = m_{K_0^*}$ ,
$X_{\sigma_{\eta '}} = m_{f_0 '}$ , $X_{\sigma_{\eta}} = m_{f_0}$).
\item Entropy density $s/T^3$, normalized energy density $u/T^4$ and
pressure $p/T^4$ vs temperature. Errors are only indicated for $s/T^3$.
\end{list}

\end{document}